\documentstyle[12pt]{article}
\newcommand{\be}{\begin{eqnarray}}
\newcommand{\ee}{\end{eqnarray}}
\begin{document}
\thispagestyle{empty}

\begin{center}
\vspace*{1cm}
 {\large{\bf Establishing the Uniqueness
of the Connection\\
between AdS$_5$ and Conformally Invariant Relativistic
Systems:\\
A Group/Field Theoretical Approach
\\}}

 \vspace*{.5cm}
{ A.~I.~Karanikas and C.~N.~Ktorides}\\
\smallskip
{\it University of Athens, Physics Department\\
Nuclear \& Particle Physics Section\\
Panepistimiopolis, Ilisia GR 15771, Athens, Greece}\\

\end{center}

\begin{abstract}
Adopting as working assumption that the conformal group O(4,2) of
Minkowski space, being the largest symmetry group which respects
its light cone structure, is the appropriate global symmetry
underlying the description of relativistic systems, it is shown
that AdS$_5$ uniquely emerges as the space on the boundary of
which a corresponding relativistic field system should be
accommodated. The basic mathematical tools employed for
establishing this result are (a) Cartan's theory of spinors and
(b) group contraction methods. Extending our considerations to
supersymmetry it is demostrated how an $N$=1 SUSY YM field system
can emerge as a broken version of an $N$=4 SUSY YM field system.
An especially important feature of the presentation is the
`unearthing' of seminal, independent from each other, works of I.
Segal and of S. Fubini which give a purely field theoretical
perspective on the intimate relation between conformally invariant
relativistic field theories and AdS$_5$ including, in particular,
the warping phenomenon.
\end{abstract}

\newpage

\section{Introductory Remarks}

 The conformal group of Minkowski
space-time is the largest symmetry group which preserves its light
cone structure. In this sense, it can be said that conformal
invariance is the maximal symmetry compatible with a
four-dimensional spacetime which does not admit absolute
simultaneity. Generally speaking, any $n$-dimensional
(pseudo)Euclidean, space $E_{m,n-m}, m\leq n$, has $O(m+1,n-m+1)$
as its corresponding conformal symmetry group\footnote{Given the
fact that spinors enter our analysis in a basic way, we shall,
throughout this exposition, refer to full orthogonal groups
instead of their simply connected components; e.g. O(3) instead of
SO(3), unless we are explicitly referring to the proper part
and/or its covering.}. A given physical system, formulated in
$E_{m,m-n}$ and so constructed as to be symmetric under
transformations induced by its conformal group, has extremely
stringent properties the most characteristic aspect of which is
that it does not allow, by definition, the introduction of any
{\it a priori} given scale(s). Conformal symmetry has proven
itself a valuable tool in specific situations such as the
theoretical analyses of scattering processes at very high
energies, but more importantly, it plays a central role in efforts
to attack fundamental theoretical issues from a global
perspective. Historically speaking, perhaps the best known
mathematical construction which admits conformal flatness is that
of Penrose [1], which is expressed in the language of {\it
twistors}. A lesser known space-time scheme that adopts conformal
invariance (in four space-time dimensions) is Segal's
chronogeometry [2], which is formulated in a more conventional
language. In our times, string theory dominates efforts aiming at
the unified description of our physical cosmos at a fundamental,
microscopic level. In this context, the game widens so that in
addition to space-time the so-called, in the old language,
internal type symmetries are included as well. The AdS/CFT
conjecture [3], in particular, relates certain conformally
invariant field systems, defined on the (four dimensional)
boundary of AdS$_5$ space, to corresponding (super)string theories
defined inside AdS$_5$, modulo an `internal' space. A primary
example is the ${\cal N}=4$ supersymmetric Yang-Mills theory, in
the large N limit, on the boundary of AdS$_5$, being dual to a
Type IIB superstring theory `living' inside AdS$_5$($\times S^5$).

The central objective in this work goes, in a sense, the oposite
way. The idea is to explore the geometrical profile of a, generic,
conformally symmetric field system in Minkowski space and
determine, constructively, how such a field system finds a natural
accommodation on the boundary of AdS$_5$. The relevant analysis
will heavily rely on group theory, especially through the
utilization of the method of group contractions. At the same time,
it will extensively employ Cartan's theory of spinors [4] as a
fundamental mathematical tool. According to Cartan, spinors, for a
given (pseudo)Euclidean space, are described by coordinates which
can be viewed, in a sense, as `square roots' of a set of
tensors\footnote{This is related to the fact that spinors are
fundamentally associated with reflections and two reflections
amount to a rotation.} of various degrees belonging to a {\it
Clifford algebra} associated with the (pseudo)Euclidean space. In
different words, the said tensors can be expressed as bilinears in
spinorial coordinates, an occurrence which, among other things,
can be used as a criterion for identifying positive definite
elements of the Clifford algebra. Utilizing these tools the
connection between AdS$_5$ and conformal field systems will emerge
in a natural manner and new insights regarding the AdS/CFT duality
conjecture will be gained.

The exposition in this paper is organized as follows. In section 2
we shall direct our thinking towards mapping a course whose
starting point is the adoption of the conformal symmetry group
O(4,2) of the Minkowski spacetime as the underlying symmetry
characterizing a given relativistic (field theoretical) system of
interest and subsequently devises a systematic reduction
procedure, guided by the following requirement: The proper
relativistic Hamiltonian, equivalently, time development operator
of the constructed system is the maximally positive generator of
the group\footnote{Clearly, this Hamiltonian should tend, for any
local measurement, to the conventional relativistic one, {\it
i.e.}, in group theoretical terms, to the Poincar\'{e} generator
$P_0$.}. As it will turn out, via the utilization of Cartan's
theory of spinors, the realization of such a goal invariably
passes through the anti de Sitter space AdS$_5$ which, once
compactified, acquires a local Minkowskian structure at its
boundary. The intermediary role of AdS$_5$ as fundamental
component of relativistic descriptions which recognizes the
necessity to distinguish between the adequacy of the Poincar\'{e}
group for `local' relativistic descriptions and a `takeover' by a
maximally positive generator of the conformal group O(4,2), at
very large distances, was recognized long time ago by Segal[2]
whose, relevant, chronogeometric theory will be discussed in
Section 3. Operating independently, in a field theoretical
context, Fubini [5] came into exactly the same realization guided
by his interest to determine the appropriate manner  by which an,
originally adopted, conformally symmetric relativistic field
system should break spontaneously in order to accommodate
realistic descriptions of physical processes associated, {\it
e.g.}, with particle masses. This approach will be considered in
Section 4. An extension to a, corresponding, supersymmetric
scenario will be subsequently presented in Section 5, where,
following Fubini's spontaneous symmetry breaking approach, an
explicit construction will be presented which demonstrates how a,
conformally invariant, $N$=4 supersymmetric Yang Mills system
naturally breaks into, for example, $N$=1 super YM one.


\section{Some Mathematical Preliminaries}

Consider some n-dimensional (pseudo)Euclidean, space $E_{m,n-m},
m\leq n$. One associates with it a $2^{n\over2}(2^{n-1\over 2})$
dimensional spinor space ${\cal S}$ for $n$ even(odd). At the same
time a corresponding {\it Clifford algebra} ${\cal C}_n$ can be
constructed whose only non-trivial, finite dimensional irreducible
representation is given in terms of $2^{n\over2}\times
2^{n\over2}\, \left(2^{n-1\over 2}\times 2^{n-1\over 2}\right)$
matrices. The $2^n$($(2^{n-1\over 2})$-dimensional algebra ${\cal
C}_n$ is so organized as to contain the unit scalar, $n$
one-vectors, $\frac{n(n-1)}{2}$ two-vectors, etc. up to and
including the unit pseudoscalar ($n$-vector). The one-vectors of
${\cal C}_n$ are in a 1-1 correspondence with the elements of a
vector base of the underlying (pseudo)Euclidean space $E_{m,n-m}$
in the spinorial representation. The 2-vectors correspond to rank
2 tensors and so on up to and including the unit pseudoscalar,
which is identified with the $n$-vector. The $2^{n\over2}\times
2^{n\over2}\, \left(2^{n-1\over 2}\times 2^{n-1\over 2}\right)$
irreducible representation of the Clifford algebra elements
establishes a common language between spinors and tensors.
Finally, the 2-vectors of ${\cal C}_n$ are in one to one
correspondence with the generators of the rotation group of the
underlying (pseudo)Euclidean space $E_{m,n-m}$ (spinorial
representation thereof).

In Ref. [6] a lemma was proved, for the particular case of the
group O(4,2), according to which given two spinors $\xi,\,\xi'$ in
${\cal S}$ there always exist bilinear forms $Y_{AB\cdot\cdot\cdot
C}(\xi,\xi')$, indices running through the values\footnote{Spinor
component indices are represented by capital Latin letters.}
0,1,2,3,5,6 and such that $A<B<\cdot\cdot\cdot<C$, which form the
components of some $p$-vector. Furthermore, it has been shown
that, if instead of two different spinors one uses components of a
single spinor $\xi$ to form bilinear expressions, one can only
form components of a 2-vector, a 3-vector and a 6-vector (unit
pseudoscalar). Of crucial importance to the proof of the lemma is
the involvement of the conjugation matrix $J$, the analogue of
$\gamma_0$ for the Minkowski case, which takes spinor $\xi$ to its
conjugate spinor $\bar{\xi}(\equiv\xi^TJ$, for real spinors). As
Cartan establishes, $J$ is a $p$-vector formed by the Clifford
product among all the 1-vectors which correspond to the
(pseudo)Euclidean directions with positive signature, {\it i.e.}
the `time' directions\footnote{Our convention for the Minkowski
metric, which appropriately adjusts to the other, higher
dimensional spaces entering our analysis, is +,-,-,-.} in physics
language. In the case of O(4,2) $J=\beta_0\beta_6\equiv J_{06}$,
where $\beta _0$ and $\beta _6$ are the Clifford 1-vectors
assigned to the 0- and 6-direction, respectively.

A result of utmost importance from Cartan is the following: Among
all elements of ${\cal C}_n$ the one which is given as a maximal,
positive definite bilinear expression in terms of spinorial
components is {\it precisely} $J$. The proof [6], basically rests
on the fact that the bilinear form $(-1)^j\xi^TJX_ {(p)}\xi$,
which applies to any p-vector $X_{(p)}$ and where $(-1)^j$ is a
phase-factor associated with the reflection of $X_{(p)}$ with
respect to a given E(4,2) vector $\vec{a}$, becomes positive
definite only when $J$ is substituted for $X_{(p)}$, as it so
happens [4,6] that $J^2=(-1)^j$ for the pseudo-Euclidean space
$E(4,2)$. This means that the maximal positive definite element
happens to be a 2-vector {\it hence a generator of the group}
$O(4,2)$. Note, in passing, that neither for $O(5,1)$ nor for
$O(3,3)$ is the maximal positive definite bilinear a generator of
the corresponding ``rotation'' group. Given that for $O(4,2)$ the
generator $J_{06}$ has a (maximally) positive definite spectrum,
we shall adopt, as a working hypothesis,\ that {\it it} should be
one's choice for representing the energy operator of a conformally
invariant system, a choice which, of course, is subject to
falsification. Let us also mention that, as it turns out (see
relevant footnote in section 5), the conventional Minkowski space
energy operator $P_0$ is also positive definite, but it is only a
{\it part} of $J_{06}$ in the sense that whereas the latter
generator is a positive definite bilinear (in spinorial
coordinates) composed of eight terms, $P_0$ is given by a subset
of only four of them. From a, local, field theoretical point of
view the motivation for testing the viability of $J_{06}$ as {\it
the} energy operator of a relativistic system will be based on
evidence, see, for example, Ref. [2], that its physical
descriptions locally coincide with those associated with $P_0$. At
very large distances, on the other hand, it takes over -being a
basic ingredient of conformal symmetry- as the proper energy
operator, equivalently, time development, generator. An
observation of note is that $J_{06}$ enters as a generator of a
homogeneous group, as opposed to $P_0$ which belongs to the
`inhomogeneous' sector of the Poincar\'{e} group.

Our findings, to this point, suggest that the energy operator
associated with conformally invariant descriptions in Minkowski
space-time should correspond to the generator of rotations with
respect to the 0-6 directions in a five dimensional `sphere'
$S_{4,2}$ specified by \be \eta_0^2-\eta^2_1-\eta^2_2-\eta^2
_3-\eta^2_5+\eta^2_6=const \ee and as such it generates the $O(2)$
factor of the maximally compact connected subgroup of O(4,2),
which is isomorphic to $O(4)\times O(2)$. Now, the rank of the
group O(4,2) is 3. This entails the presence of three Casimir
operators. Consistency with the, local, Minkowski space
instruction that elementary particle entities need two Casimir
operators for their full specification calls for a reduction from
O(4,2) to a rank two subgroup which properly characterizes
particle entities in a given local measurement. To this end, we
shall enlist the aid of the method of group contractions from a
(pseudo)orthogonal O(p,q) to an inhomogeneous (pseudo)orthogonal
group acting on a space with one less homogeneous dimension.
Specifically, one has \be O(p,q) \rightarrow IO(p,q-1)\,\,{\rm
or}\, IO(p-1,q),\quad p\not= q. \ee

We recall that the process of contraction has the following
picture. Given an O(p,q)-invariant hypersphere one imagines a
locally perpendicular patch to a given direction, stretching to
infinity so that the whole, O(p,q)-invariant configuration tends
to a (p+q-1)-dimensional flat space with IO(p,q-1), or IO(p-1,q),
its group of isometries. The particular outcome depends on the
orientation of the patch  being stretched. In this way some of the
rotation generators become translational ones. In the present case
the requirement that the generator $J_{06}$ remains intact
uniquely points to the contraction $O(4,2)\rightarrow IO(3,2)$. It
will be now demonstrated that this commitment will be realized
once the constant appearing on the rhs of Eq. (1) has a negative
value, {\it i.e.} the original O(4,2)-invariant hypersphere is
given by \be
\eta_0^2-\eta_1^2-\eta_2^2-\eta_3^2-\eta_5^2+\eta_6^2=-R^2. \ee To
this end, let us recall that the contraction process results
through the following procedure. One considers standing in the
vicinity of the ``north pole'', (0,\,0,\,0,\,0,\,R,\,0), where
$\eta_5$ has been chosen as the ``north'' direction\footnote{The
patch is locally tangential to the pole.}.

 Given, now, the generators $J_{AB}=i\left(g_{AC}\,
\eta^C\frac{\partial}{\partial\eta^B}-g_{BC}\,\eta^C\frac{\partial}{\partial\eta^BA}\right)$
of O(4,2) one redefines them by setting
$Y_{\alpha\beta}=J_{\alpha\beta}$, if $J_{\alpha\beta}$ does not
involve $\eta _5$ and $P_\alpha={1\over R}J_{\alpha\beta}$, with
$\alpha=0,1,2,3,6$ and $\beta=5$. Upon taking the limit
$R\rightarrow\infty$, the hypersurface tends towards a flat,
5-dimensional pseudo-Euclidean space (AdS$_5$) and one obtains the
algebra of IO(3,2):
\begin{eqnarray}
&&[Y_{\alpha\beta},Y_{\gamma\delta}]=i\{g_{\alpha\delta}Y_{\beta\gamma}
-g_{\alpha\gamma}Y_{\beta\delta}+g_{\beta\gamma}Y_{\alpha\delta}-g_{\beta\delta}Y_{\alpha\gamma}\}
\nonumber\\ && [P_\alpha,\,P_{\beta}]=0\nonumber\\
&&[Y_{\alpha\beta},\,P_\gamma]=i\{g_{\beta\gamma}P_\alpha-g_{\alpha\gamma}P_\beta\},
\end{eqnarray}
with $g_{\alpha\beta}={\rm diag}(+,-,-,-,+)$. In other words, by
having set $const=-R^2$ in Eq. (1) it has been ascertained that
one of the negative signature directions, $\eta_5$ in our case,
has been eliminated\footnote{In the sense that the group
contraction process reduces the homogeneous dimension of the
space-time manifold by one, while compensating via the
introduction of space-time translation generators.}. Finally, note
should be taken of the fact that mathematical consistency requires
that the O(4,2)-invariant sphere should, in its Euclidean version,
be `punctured' at a point, e.g. `` south pole'' in order for our
construction to achieve the asymptotic flatness.

 The homogeneous part of the contracted group acts naturally on an
SO(3,2)-invariant, four-dimensional `hypersphere' $S_{3,2}$. The
latter can be projected onto the original O(4,2)-invariant
`hypersphere' $S_{4,2}$ anywhere on a locus which will appear as a
``trajectory'' of $S_{3,2}$ in $S_{4,2}$\footnote{In the sense of
the Euclidean analogue of $S_n/S_{n-1}\simeq S_1$}.

The inhomogenous part of the contraction pertains to translational
generators in AdS$_5$. Of utmost importance is the fact that the
rank-2 symmetry subgroup has two Casimir operators and contains
$J_{06}$ as one of its generators. The original O(4,2) symmetry,
is expected to still be operational in one way or other and this
matter will draw a considerable portion of our attention
throughout this work.

For now let us make a first connection with the AdS/CFT duality
scenario according to which the conformal field theory component
`lives' on the, four dimensional, {\it boundary} of AdS$_5$.
Accordingly, a compactification procedure is called for. This
matter has been given special attention by Witten in [7] (see also
[8]) who, working in Euclidean formalism, has dealt with the issue
by adding a point at infinity which accomplishes the task. This
act, in the presently advocated scheme, we interpret as `putting
back' the point that was `taken out' during the contraction
procedure. Finally, the inhomogeneous part of IO(3,2) commutation
relations refer, for the compact version of AdS$_5$, to
translations in the interior of the ball, while O(3,2) is
associated with rotations of a five-dimensional sphere. Finally,
the local Minkowski character of $S_{3,2}$ emerges through the
contraction $O(3,2)\rightarrow IO(3,1)$. This implies that in the
flat limit, which is equivalent to saying `locally', our universe
becomes Minkowski space and our geometrical group contracts to
that of Poincar\'{e}\footnote{In fact, the contraction
$O(3,2)\rightarrow IO(3,1)$ involves the mapping of one of the
generators $J_{ab}$ with $b=0$ or 6 into the translation operators
$P_\mu$.}.

In closing this section let us make a quantitative remark relating
the O(3,2)-invariant hypersphere to the O(4,2)-invariant one with
which we started. As already pointed out, for a fixed value of
$\mid R^2\mid$ an O(3,2)-invariant hypersphere $S_4$, in Euclidean
version, can be placed anywhere on a one dimensional circular
trajectory in $S_5$. A given choice, of course, fixes a specific
4-dimensional `sphere' $S_{3,2}$. Let, now, $r'$ be the radius of
the aforementioned `trajectory' corresponding to the case where
the `pole point' $\eta_5$ is inserted on the rhs of (1), while $r$
the respective radius when $R(<\eta_5)$ is inserted. We then have
that ${r\over r'}={R\over \eta_5}$. Substituting into the equation
which defines the hypersphere for the arbitrary value of $\eta_5$
one obtains
\begin{equation}
\eta_0^2-\eta_1^2-\eta_2^2-\eta_3^2+\eta_6^2=-R^2+\frac{r'^2}{r^2}R^2.
\end{equation}

Introducing the set of $\zeta$-coordinates, where
$\zeta_a=\eta_a{r\over r'}$ one writes
\begin{equation}
\zeta_0^2-\zeta_1^2-\zeta_2^2-\zeta_3^2+\zeta_6^2=R^2\left(1-\frac{r'^2}{r^2}\right)\equiv
a^2.
\end{equation}
Notice that the positive definite character of $a^2$, which can be
surmised from the fact that $r<r'$ and ${\rm sign} r={\rm sign}\,
r'$, confirms that it is one of the directions 0 or 6 which
`flattens up' in the limit.


\section{Segal's chronogeometry}

Our discussion, to this point, has followed a general line of
reasoning, which promoted a mathematical scenario according to
which it is the conformal group of Minkowski space that describes
globally, the spacetime symmetries of our world, while it is
expected to approach Poincar\'{e} group based descriptions at
local level. Our immediate obligation is to demonstrate that the
differences between $J_{06}$ and $P_0$ are unobservably small for
sufficiently local Minkowskian  regions. To this end, we now turn
our attention to Segal's chronogeometry [2], which was developed
by the author for, among other things, `rationalizing'
observational data regarding motions of stars at extragalactic
distances. The, expected, significant departures between $J_{06}$-
and $P_0$-based estimations of the velocity of distant stars turns
out to overwhelmingly favor the former over the latter.

Following Segal we associate the generator of time development
with that of rotations in the 0-6 plane. The corresponding `time'
parameter $\tau$ is thereby identified with the angle of such
rotations. One writes
\begin{equation}
\frac{\xi_0}{\xi_6}=\tan\tau,
\end{equation}
with $-\pi<\tau<\pi$, a periodicity which brings to surface the
well known problem regarding conformal invariance and causality.
Its confrontation calls for reverting to the universal covering of
the proper group  SO(4,2), namely SU(2,2). In doing so the one
parameter subgroup $\{T_t\}$ generated by $J_{06}$ is covered an
infinite number of times, equivalently, the SU(2,2) chronometric
world becomes an infinite-sheeted four dimensional manifold
$\tilde{M}$. A point on $\tilde{M}$ is described by a set of
coordinates $(\tau,u_1,u_2,u_3,u_4)$, where $u=(u_1,u_2,u_3,u_4)$
is a point on a four-dimensional Euclidean sphere, a specification
implied by the, local projective identification of the
SO(3,2)-invariant hypersphere and the Minkowski space. Explicitly,
the projective identification of the five-dimensional coordinates
$(\zeta_0,\zeta_1,\zeta_2,\zeta_3,\zeta_6)$, transforming like the
components of an SO(3,2)-vector and the Minkowski coordinates is
given by the relations
\begin{equation}
\zeta_\mu=\frac{2a^2x_\mu}{a^2+x^2},\quad
\zeta_6=\frac{a(a^2-x^2)}{a^2+x^2},
\end{equation}
where $a$ is a fixed quantity with the dimension of length. It
immediately follows that
\begin{equation}
\tan\tau=\frac{ax_0}{a^2-x^2}.
\end{equation}

Upon introducing
\begin{equation}
u_j=\frac{a \zeta_j}
    {\left[\zeta^2_0+\zeta^2_6\right]^{1\over 2}},\quad j=1,2,3
\end{equation}
and
\begin{equation}
u_4=\frac{a^2}
    {\left[\zeta^2_0+\zeta^2_6\right]^{1\over 2}},\quad j=1,2,3,
\end{equation}
one immediately obtains that $\sum\limits_{j=1}^{4}u_j^2=a^2$,
{\it i.e.} $a$ is the radius of a 4-dimensional {\it Euclidean}
sphere\footnote{Corresponding to the four spacelike directions.}.
Furthermore, one obtains the relations
\begin{equation}
u_j=2\lambda x_j,\quad u_4=\frac{\lambda(a^2+x^2)}{a}\, ,
\end{equation}
where
\begin{equation}
\lambda=\frac{a^2}
 {[(a^2-x^2)^2+4a^2x_0^2]^{{1\over 2}}},
 \end{equation}
 with $x^2=x_0^2-x_1^2-x_2^2-x_3^2$.

In this way one is able to relate the Minkowski coordinates to
Segal's chronometric ones $(\tau,u_j)$ on the, 4-dimensional
O(3,2)-invariant, hypersphere, {\it i.e.} AdS$_5$ space. More
precisely, once the measure of $a$ is set to unity, the above
relations give Segal's mapping effecting the embedding of the
Minkowski space in $\tilde{M}$. The interested reader regarding
basic issues such as causality, simultaneity, quantization,
masses, {\it etc.} the is referred to Segal's papers. As a
specific example we here outline Segal's derivation of a ``red
shift phenomenon'' associated with distant observations in the
``chronometric universe'' $\tilde{M}$ [2]. One starts by observing
that the `time displacement' (dually energy) operator
$J_{06}\equiv H$ can be split into two parts, $H=H_0+H_1$. As it
turns out, $H_0$, is scale covariant, while the second is
anti-scale covariant and respectively identify with the generators
of $P_0$ and $K_0$, {\it i.e.} the zero components of the
translation and special conformal transformations. Each one is
given as a positive definite quantity, {\it i.e.} as sum of four
square terms (in spinorial coordinates). Between them they share
the eight terms entering the expression for $J_{06}$.

As it turns out, for local events, with respect to a given
observer, the effects `evolving' through $H_1$ are negligible. For
large distances, on the other hand, there arise notable
differences. Suppose, for example, that a photon has been emitted
from a distant star. Its energy will be measured locally and at
the time of its emission is determined by the operator
$\hat{H}_0$. The development in chronometric time is given, in the
Heisenberg picture, by
\begin{equation}
H_0(\tau)=e^{-i\hat{H}\tau}H_0e^{i\hat{H}\tau}.
\end{equation}
Given the non vanishing of the commutator $[H,H_0]$, one writes,
group theoretically,
\begin{equation}
H_0(\tau)=\alpha H_0+\beta H_1+\gamma[H_0,H_1],
\end{equation}
where $\alpha,\,\beta$ and $\gamma$ are functions of $\tau$. A red
shift factor $Z$ emerges once the expectation value of $H_0(\tau)$
is compared with that of $H_0$. The following result is obtained
[2]
\begin{equation}
\langle H_0(\tau)\rangle={1\over 1+Z}\langle H_\rangle,
\end{equation}
where $Z=\tan^2{\tau\over 2}$.

 Plotting $\log Z$ against
cosmographical parameters, Segal obtains remarkable agreements
with existing data. A similar red shift factor, going by the name
of `warping', rises in connection with the AdS/CFT duality scheme
[8].

\section{Dynamical aspects: Fubini's field theoretical approach}

The theoretical considerations developed in the previous section
are, basically, of `geometrical' nature. In this section we shall
gain an alternative perspective on the general theme we have been
consistently developing in the preceding considerations by taking
a point of view which focuses on dynamical implications. In
particular, we shall proceed to assess the viability of
spontaneous symmetry breaking mechanisms operating on an,
originally, conformally invariant field theoretical equations
whose solutions exhibit a breakdown to conventional relativistic
form, beyond a given energy regime. To this end we follow Fubini
[5] by considering a (simple) system defined by the Lagrangian
density
\begin{equation}
{\cal L}=-{1\over 2}\partial_\mu\phi\partial^\mu\phi-g\phi^4,
\end{equation}
which does not contain any dimensional parameter. We are
interested in exploring nontrivial solutions for this system of
the form
\begin{equation}
\phi(x)=B(x)+\phi'(x),
\end{equation}
where $B(x)$ is a classical solution of the field equation and
$\phi'(x)$ a small quantum disturbance such that $\langle
0\mid\phi'(x)\mid 0\rangle =0$. Normalizing the vacuum state to
unity we have
\begin{equation}
\langle 0\mid\phi'(x)\mid 0\rangle =B(x).
\end{equation}

Now, being a classical solution, $B(x)$ satisfies the conformally
invariant equation
\begin{equation}
\partial^\mu\partial_\mu B+4gB^3=0.
\end{equation}
In a search for particular, conformally invariant, solutions one
is guided by symmetry considerations for the ground state.
Specifically, given a generator $G_\kappa$, of the conformal
group, expressed in differential form, a solution of (20) will be
invariant under the action of $G_\kappa$, if
\begin{equation}
\langle 0\mid[G_\kappa, B(x)]\mid 0\rangle=0
\end{equation}
holds true. One expects that only the trivial solution $B(x)=0$
satisfies the above equation, if $G_\kappa$ runs through all
generators of O(4,2). If one requires that the invariance is with
respect to all the Poincar\'{e} group generators, but not any of
the rest, then the solution $B(x)=$const. is the most general one,
as can be demonstrated by the action under $P_\mu$, {\it i.e.}
\begin{equation}
i\frac{\partial B(x)}{\partial x^\mu}=0,
\end{equation}
coinciding, as expected, with the solution of the free equation
$(g=0)$.

Consider, now, the case where one demands invariance under the
action of the generators $R_\mu={1\over 2}\left(aP_\mu+{1\over
a}K_\mu\right)$, where $a$ enters as a fundamental length,
necessary for balancing the units of the two terms entering the
sum\footnote{Obviously the dimensional analysis pertains to the
interpretation of the various operators in the context of
Minkowski space.} . It leads to the equation
\begin{equation}
\frac{a^2+x^2}{2}\,\frac{\partial B(x^2)}{\partial x^\mu}+x_\mu
B(x^2)=0,
\end{equation}
with the argument $x^2$ serving to take account of the fact that
the Lorentz `rotation' symmetry remains intact.

The solution of Eq. (22) is given by [5]
\begin{equation}
B(x^2)=\frac{1}{\sqrt{2g}}\frac{a}{x^2+a^2}.
\end{equation}

In a set of six dimensional coordinates appropriate to $\tilde{M}$
and associated with the covering group $SU(2,2)$, defined, in
obvious notation, by $v_iv^i=-2g,\quad v_\mu=u_\mu,\quad
u_5=\frac{1-x^2}{2},\quad u_6=\frac{1+x^2}{2}$, the above solution
takes the simple form
\begin{equation}
B(u)=(v_i u^i)^{-1}.
\end{equation}

It turns out, that the six-vector $v_i$ indicates the direction
along which the O(4,2) symmetry breaks during the contraction to
IO(3,2), {\it e.g.} the $\eta_5$-direction for the procedure
adopted in Section 2. One actually verifies [5] that for a
positive value of $g$ the six-vector $v_i$ lies on the hyperboloid
$v_iv^i=-R^2$, respectively negative sign for the contraction to
IO(4,1) (de Sitter space). A different way of assessing the
situation we have just analyzed is to say that the breaking of the
conformal symmetry towards the anti-deSitter {\it vs.} deSitter
direction depends on the sign of the coupling constant $g$.

It will now be demonstrated that the O(4,2) generator $J_{06}$
coincides, via the contraction process, with the ${1\over
2}\left(aP_0+{1\over a}K_0\right)$ combination of IO(3,2)
generators. To this end consider the relations in Eq. (8) which
relate the Minkowski coordinates to the five coordinates $\zeta_i$
for the O(4,2)-invariant sphere. In terms of the latter one writes
\begin{equation}
J_{06}=i\left(\zeta_0\frac{\partial}{\partial
\zeta_6}-\zeta_6\frac{\partial}{\partial \zeta_0}\right),
\end{equation}
which yields
\begin{equation}
J_{06}={i\over 2}\left[\left(a-{x^2\over
a}\right)\partial_0+2\frac{x_0x^\nu}{a}\partial_\nu\right].
\end{equation}
This clearly coincides with ${1\over 2}\left(aP_0+{1\over
a}K_0\right)$. It can be similarly shown that $J_{6\mu}={1\over
2}\left(aP_\mu+{1\over a}K_\mu\right)$, while the remaining
generators of O(3,2) coincide with the Lorentz ones, {\it i.e.}
$M_{\mu\nu}$.

It might be of interest, at this point to reproduce an argument by
Segal [2], which confirms the positive definiteness of $J_{06}$:
As is well known, $K_\mu={\cal I}P_\mu{\cal I}$, where ${\cal I}$
is the inversion operator, which induces the, Minkowski space
transformation $x_\mu\rightarrow \frac{x_\mu}{x^2}$. on
$\tilde{M}$ space, where it appears as the singularity free
transformation $(\tau,u)\rightarrow (\pi-\tau,u)$. Hence, ${\cal
I}$ is continuously connected with the time reversal
transformation, {\it i.e.} it is represented by an antiunitary
operator. Thus, ${\cal I}P_0{\cal I}$ has a positive spectrum when
$P_0$ does.

The energy density in Fubini's scheme is given by the expression
\begin{equation}
{\cal E}={1\over 2}\left(aT_{00}+{1\over a}K_{00}\right),
\end{equation}
where $T_{\mu\nu}$ is the energy momentum tensor and $K_{\mu\nu}$
a local tensor current associated with the conformal charges. For
the particular model under consideration it is given by
\begin{equation}
K_{\mu\nu}=2x^\rho x_\nu T_{\mu\rho}-x^2
T_{\mu\nu}+2x_\nu(\partial_\mu\phi)\phi.
\end{equation}

It is a straightforward task to calculate ${\cal E}$ for the
classical solution given by (23). One obtains
\begin{equation}
{\cal E}=\frac{2a(x_E^2+a^2)(a^2-x^2)}{g(a^2+x^2)^4},
\end{equation}
where $x_E^2$ denotes the Euclidean magnitude
$x^2_0+x^2_1+x^2_2+1x^2_2+x^2_3$.

The corresponding relativistic expression, {\it i.e.} the one
formulated in Minkowski space, gives
\begin{equation}
{\cal E'}=\frac{2a^3(x_E^2+a^2)}{g(a^2+x^2)^4}.
\end{equation}
The difference $\Delta{\cal E}={\cal E}-{\cal E'}$ is seen ro be
positive on a space-like surface, another verification of the
maximality of the energy associated with this approach. This red
shift effect reproduces, once again the warping associated with
the AdS/CFT duality scheme.

\section{Supersymmetry Considerations}

In this section we extend our considerations to supersymmetry. To
begin, let us recall that the original version of supersymmetry
[9] contained the conformal algebra of SO(4,2) as an integral
part, along with eight spinorial charges. We shall refer to this
as the Wess-Zumino algebra and denote it by ${\cal W}$. The
particular subalgebra of ${\cal W}$ which contains the
Poincar\'{e} generators and only four spinorial charges was
originally proposed by Volkov and Akulov [10] and will be denoted
by ${\cal V}$. In the framework of the basic theme of this work we
shall proceed to investigate possible advantages of the former
over the latter. Now, in Ref [6] it was shown that the conformal
algebra has a unique extension to the ${\cal W}$ supersymmetry, an
extension which does not seem to hold between the Poincar\'{e}
algebra and ${\cal V}$. Moreover, according to Haag {\it et al}
[11], see also Ref. [6], it is only within the framework of ${\cal
W}$ that it becomes possible to intertwine internal-type
symmetries (R-symmetries in current language) with supersymmetry
in a non-trivial way. A complete listing of all the supersymmetry
algebras can be found in the classic work of Nahm [12].

Let us recall that ${\cal W}$ is a 24-generator graded algebra
spanned by the set of generators
$\{K_\mu(+2),\,Q_\alpha(+1),\,M_{\mu\nu},\,D,\,\Pi(0),\,Q_\alpha^0(-1),\,P_\mu(-2)\}$,
where the numbers in parentheses give corresponding grades. The
immediate question is whether ${\cal W}$ can be reorganized so as
to define, along with the generators
$\{M_{\mu\nu},\,R_{\mu}=\frac{1}{2}\left(aP_\mu+{1\over
a}K_\mu\right)\}$, a self-consistent algebraic structure. To this
end, we introduce a new set of spinorial charges, $\Xi_\alpha$,
given by
\begin{equation}
\Xi_\alpha={1\over 2}\left(\sqrt{a} Q^0_\alpha
+{1\over\sqrt{a}}Q^1_\alpha\right).
\end{equation}
It is a matter of simple algebra to show that
\begin{eqnarray}
\quad&&[\Xi_\alpha,R_\mu]=(\gamma_\mu)_\alpha
^\beta\Xi_\beta\nonumber\\&&
\{\Xi_\alpha,\Xi_{\beta}\}=\left[(\gamma^\mu\gamma_0)_{\alpha\beta}R_\mu
-{1\over
2}(\gamma^{\mu\nu}\gamma_0)_{\alpha\beta}M_{\mu\nu}\right],
\end{eqnarray}
with the $\gamma$'s in the Majorana representation. Consequently,
a graded algebra, to be referred to as ${\cal Z}$, spanned by the
set $\{M_{\mu\nu},\,R_\mu,\,\Xi_\alpha\}$ is formed, which is a
subalgebra of ${\cal W}$, non-isomorphic to ${\cal V}$. This
supersymmetric algebra has O(3,2) as its spacetime componenent and
admits, accoring to Nahm's classification $o(N),
\,N=1,2,\cdot\cdot\cdot,$ as its `internal' symmetry algebra. In
connection with QCD and taking into account the analysis of
Polchinski and Strassler [13], one scenario of considerable
interest arises from the breaking of the, $N$=4, conformal super
Yang Mills system to a corresponding $N$=1 supersymmetric one at
some scale slightly above $\Lambda_{QCD}$. Such a scheme is
naturally accommodated by the O(4,2)$\rightarrow$O(3,2)
supersymmetry breaking procedure advocated in this section.

To relate this construction to the chronogeometric scheme one
observes that\footnote{In terms of the spinorial charges
$Q^\alpha_0$ and $Q^\alpha_1$ of {\cal W}, $P_0$ and $K_0$ acquire
the form
\[
P_0=1/8\,{\rm Tr} \left[\left(\mid Q^0\rangle\langle
Q^0\mid\right)_{\alpha\beta}+\left(\mid Q^0\rangle\langle
Q^0\mid\right)_{\beta\alpha}\right]
\]
\[
K_0=1/8\,{\rm Tr} \left[\left(\mid Q^1\rangle\langle
Q^1\mid\right)_{\alpha\beta}+\left(\mid Q^1\rangle\langle
Q^1\mid\right)_{\beta\alpha}\right].
\]}

\begin{equation}
H=1/2(aP_0+{1\over a}K_0)={1\over
4}\left[\sum\limits_{\alpha=1}^{4}\Xi^2-{1\over
4}\sum\limits_{\alpha=1}^{4}\{Q_\alpha^0,\,Q_\alpha^1\}\right].
\end{equation}

But
\begin{equation}
\{Q^0_\alpha,
Q^1_\beta\}=-2[(\gamma^{\mu\nu}\gamma_0)_{\alpha\beta}M_{\mu\nu}-(\gamma_0)_{\alpha\beta}
D+(\gamma_5\gamma_0)_{\alpha\beta}\Pi].
\end{equation}
However, in the Majorana representation $\gamma^{\mu\nu}\gamma_0$
is symmetric whereas $\gamma_0$ and $\gamma_5$ and $\gamma_0$ are
antisymmetric. Hence
\begin{equation}
\{Q^0_\alpha,
Q^1_\alpha\}=-2[(\gamma^{\mu\nu}\gamma_0)_{\alpha\alpha}M_{\mu\nu}.
\end{equation}
Consequently, the vanishing of the vacuum expectation value of $H$
leads to
\begin{equation}
0=\langle 0\mid\sum\limits_{\alpha=1}^{4}\Xi_\alpha^2+{1\over
2}{\rm Tr} (\gamma^{\mu\nu}\gamma_0)M_{\mu\nu}\mid 0\rangle.
\end{equation}

Imposing Lorentz invariance on the theory, {\it i.e.} setting
$M_{\mu\nu}\mid 0\rangle=0$, one obtains $\langle
0\mid\Xi_\alpha^2\mid 0\rangle=0$, which implies that
$\Xi_\alpha\mid 0\rangle=0$, {\it i.e.} all elements of ${\cal Z}$
annihilate the vacuum. This conclusion, on the other hand, does
not necessarily imply imply that $Q_\alpha^0\mid 0\rangle=0$. If,
in fact, it did, then the theory would be symmetric under the
whole of ${\cal W}$, an occurrence which goes against the pattern
that has been established so far, {\it i.e.} by the
(non-supersymmetric) physical implications of the Segal/Fubini
schemes.

\section{Summary and Comments}

The analysis carried out in this paper has explored the
relationship between the AdS$_5$ space and conformal descriptions
of relativistic quantum field system. The unique way by which the
anti-de Sitter space emerged, from different perspectives, by
utilizing group contraction strategies, as well as enlisting the
aid of Cartan's theory of spinors, establishes in a unique manner
the connection between AdS$_5$ and conformally invariant field
systems thereby lending further credibility to the AdS/CFT
conjecture. From a general viewpoint one might further assess the
situation by asking oneself whether the length $a$, or its inverse
(momentum), associated with the breaking of conformal invariance,
for the non-supersymmetric and -more importantly- the
supersymmetric versions) open the way for realistic applications.
In fact, if conformal invariance is only {\it spontaneously}
broken, then it remains in the background as an overall
fundamental symmetry of a given system (in reality it is simply
hidden) degenerating to Poincar\`{e}-based descriptions, {\it in a
limiting way}, when it comes to local observations. Especially
significant, in this respect, is the analysis conducted by
Polchinski and Strassler [13] which pertains to, dynamical, QCD
processes.

{\bf Acknowledgement}: It is with sincere  appreciation and
fondness to acknowledge the contribution of M. Daniel to this
work. In fact, a good portion of it was conducted in mutual
collaboration with one of the present authors (C.N.K.) which was
recorded in an unpublished paper of 30+ years ago. Unfortunately,
he declined our request to undertake jointly the project of
bringing the old paper in tune with the AdS/CFT developments,
citing the shift of his intellectual interest towards other
directions.

\end{document}